\documentclass[12pt]{article}

\usepackage{newtxtext,newtxmath}

\usepackage{graphicx}

\usepackage[letterpaper,margin=1in]{geometry}

\usepackage{xcolor}

\renewenvironment{abstract}
	{\quotation}
	{\endquotation}

\date{}

\makeatletter
\renewcommand{\fnum@figure}{\textbf{Figure \thefigure}}
\renewcommand{\fnum@table}{\textbf{Table \thetable}}
\makeatother

\usepackage{scicite}

\usepackage{url}

\usepackage{tikz}
\usetikzlibrary{positioning, shapes,arrows.meta, decorations.pathreplacing}
\usepackage{bm}

\def\scititle{A Framework for Hybrid Physics-AI Coupled Ocean Models}
\title{\bfseries \boldmath \scititle}

\author{
	L.~Zanna$^{1,9\ast}$,\and 
       W.~Gregory$^{2}$,\and
        P.~Perezhogin$^{1}$,\and    
        A.~Sane$^{2}$,\and  
        C.~Zhang$^{2,8}$,\and 
        A.~Adcroft$^{2}$,\and
	M.~Bushuk$^{2}$,\and
        C.~Fernandez-Granda$^{1,9}$,\and
        B.~Reichl$^{2}$,\and        
        D.~Balwada$^{3}$,\and
        J.~Busecke$^{3}$,\and   
        W.~Chapman$^{5}$,\and
        A.~Connolly$^{7}$,\and
        D.~Du$^{2}$,\and
        K.~Everard$^{1}$,\and
        F.~Falasca$^{1}$,\and
        R. Falga$^{1}$, \and
        D.~Kamm$^{6}$,\and
	E. ~Meunier$^{6}$,\and
        Q.~Liu$^{1}$,\and
        A.-A.~Nasser$^{2}$,\and
        M.~Pudig$^{1}$,\and
        A.~Shao$^{2}$,\and
        J.L.~Simpson$^{7}$,\and
        L.~Vogt$^{1}$,\and
        J.~Wu$^{1}$, \\	
	\small$^{1}$Courant Institute of Mathematical Sciences, New York University, New York 10012, USA.\and
	\small$^{2}$Princeton University Atmospheric and Oceanic Sciences Program, Princeton, USA .\and 
    \small$^{3}$Lamont-Doherty Earth Observatory at Columbia University, Palisades, USA.\and
	\small$^{4}$NOAA Geophysical Fluid Dynamics Laboratory, Princeton, USA.\and
    \small$^{5}$NSF-National Center for Atmospheric Research, Boulder, CO, USA.\and
    \small$^{6}$Sorbonne Université-CNRS-IRD-MNHN, LOCEAN Laboratory, Paris, France.\and 
    \small$^{7}$Columbia University, New York, NY, USA.\and 
    \small$^{8}$Department of Civil and Environmental Engineering, Rowan University, Glassboro, NJ, USA.\and 
    \small$^{9}$Center for Data Science, New York University, New York 10011, USA.\and
	\small$^\ast$Corresponding author. Email: laure.zanna@nyu.edu}
    
\begin{document} 

\maketitle

\begin{abstract} \bfseries \boldmath

Climate simulations, at all grid resolutions, rely on approximations that encapsulate the forcing due to unresolved processes on resolved variables, known as parameterizations. 
Parameterizations often lead to inaccuracies in climate models, with significant biases in the physics of key climate phenomena. 
Advances in artificial intelligence (AI) are now directly enabling the learning of unresolved processes from data to improve the physics of climate simulations. 
Here, we introduce a flexible framework for developing and implementing physics- and scale-aware machine learning parameterizations within climate models. 
We focus on the ocean and sea-ice components of a state-of-the-art climate model by implementing a spectrum of data-driven parameterizations, ranging from complex deep learning models to more interpretable equation-based models. 
Our results showcase the viability of AI-driven parameterizations in operational models, advancing the capabilities of a new generation of hybrid simulations, and include prototypes of fully coupled atmosphere-ocean-sea-ice hybrid simulations. 
The tools developed are open source, accessible, and available to all. 
\end{abstract}

\noindent

Numerical modeling has long been a cornerstone for understanding Earth's climate system. 
However, one significant challenge persists: limited computing power inhibits our ability to capture the multiscale nature of the climate system and degrades our simulations and predictions of Earth's climate. 
Existing computing power imposes a limit on the spatial resolution of climate models. 
As a result, processes with spatial scales smaller than the model grid spacing cannot be explicitly resolved. These subgrid scale processes include both vertical processes (e.g., convection, boundary layer turbulence, vertical mixing, double diffusion, cloud microphysics, and radiative transfer) and horizontal processes (e.g., mesoscale and submesoscale eddies, isopycnal mixing, and gravity wave drag). 
These phenomena span a wide range of scales (from turbulent motions just meters across to eddies stretching over 100 kilometers), impacting large-scale transport in the ocean and atmosphere, with a cascading effect on weather and climate across spatiotemporal scales. 
Some processes, particularly mesoscale ocean eddies, are now being partially resolved in high-resolution or GPU-accelerated models \cite{adcroft2019gfdl,griffies2015impacts,silvestri2024new}. 
However, fine-scale processes, such as turbulence, convection, and cloud microphysics, remain far beyond the resolution limit. 
Thus, despite growing computational capabilities, accounting for the impact of these so-called subgrid processes on the resolved scales---known as the parameterization or closure problem---remains critical for adequately representing the climate system. 

Traditionally, physics-based parameterizations have been developed using theory and a small subset of experimental or observational data to tune their associated parameters. 
In recent years, the rise of artificial intelligence (AI) and machine learning (ML) has offered new avenues for addressing the parameterization problem. While sparse observational datasets are often not ideal for ML, high-resolution simulations can generate robust datasets for learning subgrid physics.
Given large, high-quality datasets that can resolve the subgrid processes of interest, one can approximately learn their impact on the resolved scales, e.g., without a priori imposing a closed-form parametric closure model.  
These data-driven approaches can improve the representation of subgrid processes in numerical models, as demonstrated in a range of  simulations \cite{rasp2018deep,Brenowitz2019,yuval2020stable,grundner2025,zanna2020data}. 
However, their impact when implemented in state-of-the-art operational climate models remains to be demonstrated, in part due to challenges related to the infrastructure needed for data-driven parameterization development and implementation, climate model code maintenance, and computational resources.


In this work, we introduce a framework for learning, implementing, and testing scale- and physics-aware data-driven parameterizations. We discuss a spectrum of data-driven parameterizations, ranging from physics-based closures with parameters informed by data to complete learning of closures in either equation or neural network form. We focus primarily on ocean turbulence parameterizations which are essential for simulating upper ocean properties on interannual to decadal timescales. However, we also learn and bias correct structural sea ice model errors, where errors are derived from data assimilation of real satellite observations. We demonstrate our flexible framework using a state-of-the-art global ocean and sea-ice model, in which we implement and test a suite of machine-learned parameterizations. These global hybrid simulations run stably and efficiently and, in many cases, display promising improvements relative to the baseline model, including in a fully coupled atmosphere-ocean-sea-ice setting. Our results demonstrate a promising potential for hybrid AI-physics climate modeling at scale.

\section*{AI for subgrid ocean physics}


Physics-based parameterizations have been mainstream in climate models for decades, offering numerous advances \cite{gent1990isopycnal}. 
Yet, they suffer from some drawbacks.  
Physics-based parameterization development relies primarily on scale separation between the unresolved and resolved scales. 
In the current generation of models, as resolution increases, the scale separation becomes weaker for specific physical processes (e.g., mesoscale eddies as discussed in the introduction). 
In addition, many existing parameterizations are not scale-aware; thus, they may be pushed outside their range of validity as model resolution increases. 
Even if the scale separation holds up, parametrizations are not well-constrained or informed by data \cite{schneider2017earth}. 
Finally, it might not be possible to parameterize every subgrid process using diagnostic equations as a function of coarse variables. 
Hence, new approaches are warranted.

AI and ML methods can automatically discover complex structures without prior knowledge or assumptions, making them powerful tools for learning subgrid physics at different target model resolutions. 
AI-based parameterizations use ML algorithms (e.g., neural networks) to learn missing physics; the core idea is to express the effect of subgrid processes (e.g., output of a neural network) as a function of the resolved variables (e.g., input of a neural network). 
For example, in turbulence modeling, the input might be the velocity field on the coarse-resolution grid, and the output could be the turbulent stress tensor obtained from filtering and/or coarse-graining high-resolution simulations \cite{christensen2022parametrization}.
ML models are trained using data by minimizing a loss function that quantifies the error between the predicted subgrid quantities and the true (target) values from a high-resolution simulation or experimental data.
One of the main downsides of this approach is that ML algorithms may not satisfy physical constraints (e.g., conservation laws, invariance, symmetries). 
Studies have shown that increasing the amount of data may help \cite{guan2022stable}, or that imposing such physical constraints is possible \cite{zanna2020data,BeuclerPRL2021}. 

\begin{figure}
    \centering
    \includegraphics[width=1\linewidth]{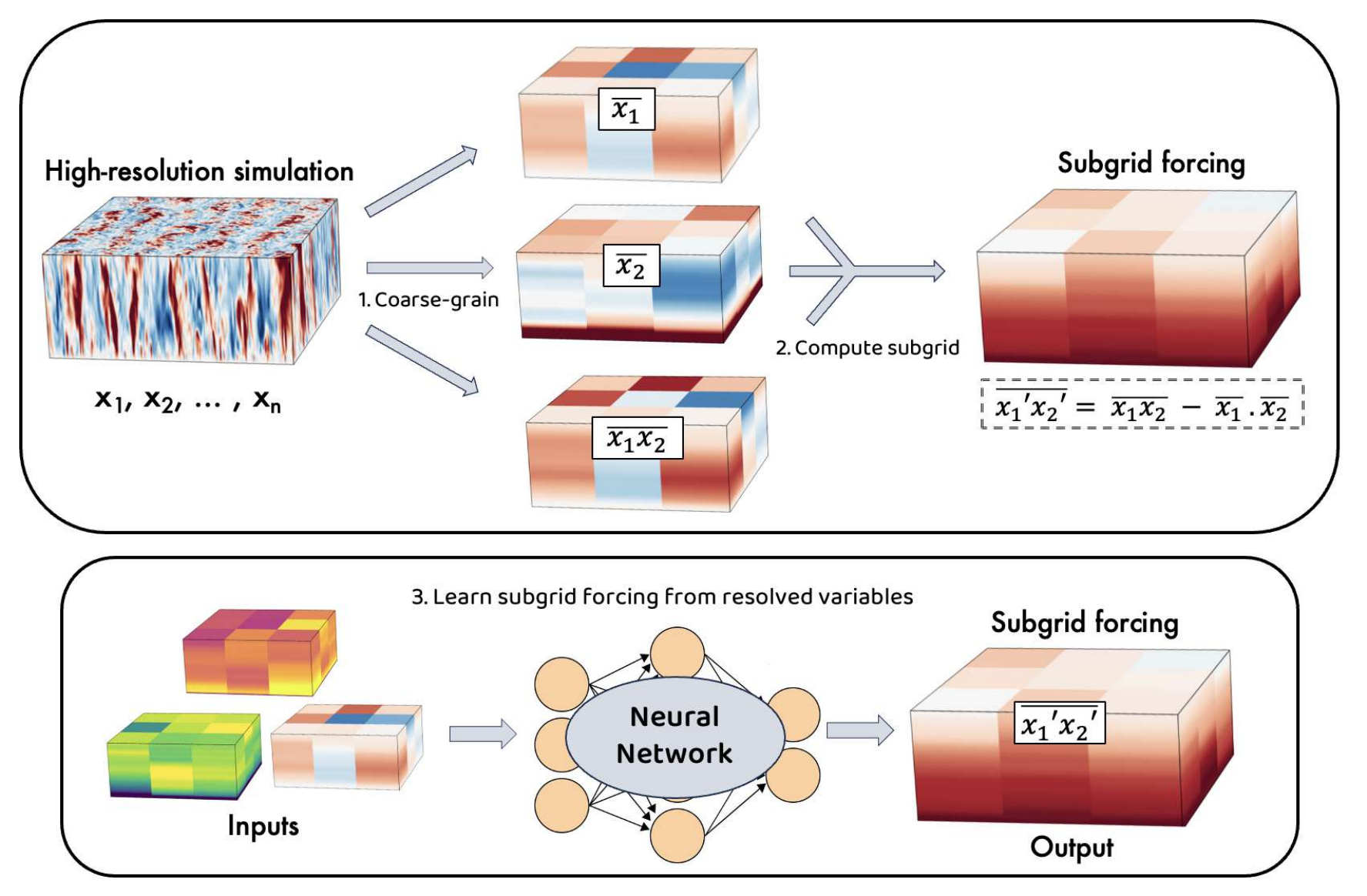}
    \caption{Schematic of learning subgrid processes from high-resolution simulation. Here, $x_n$ represents the physical variables (e.g., velocities at a grid point) simulated by a high-resolution model. A coarse-graining procedure, which relies on spatial filtering, is applied to the high-resolution data to explicitly compute the subgrid forcing and input variables on a coarse-resolution mesh. The overbar represents a spatial average, and the prime represents the fluctuation from the mean. A neural network is then trained to predict the subgrid processes (e.g., $\overline{x'_1 x'_2}$) from coarse-resolution variables (e.g., $\overline{x_1}$).
}
    \label{fig:subgrid}
\end{figure}

High-fidelity simulations, which directly resolve the physical processes to be parameterized at high spatial and temporal resolutions, are the primary source of data for learning data-driven subgrid physics in data-poor fluid problems, such as the ocean.   
A training set with low-resolution data and corresponding ground truth of the subgrid processes is built by filtering and coarse-graining the high-resolution model data (see Figure \ref{fig:subgrid}).
Such high-resolution simulations can be run over short timespans, as they focus on fast processes rather than long timescales; this can be leveraged to produce a large range of datasets with different forcings under different dynamical regimes.
There are obvious drawbacks to this approach. 
The simulations are still discretizations of the equations of motion, which can be affected by other subgrid closures (e.g., the Smagorinsky subgrid model in Large Eddy Simulations). 
In addition, high-resolution simulations typically focus on resolving a single process, thereby lacking interactions with other scales of interest. 
Despite these caveats, high-resolution and/or process-based simulations are well-established tools for understanding and parameterizing turbulence, in particular for ocean processes \cite{jansen2015parameterization, wagner2025formulation}, as actual observations are often too sparse and noisy to be the primary means of developing subgrid models. 
Although here we primarily discuss the use of high-fidelity datasets for improving subgrid parameterizations themselves, a global network of observational data also serves as valuable large-scale constraints during the parameter tuning process \cite{schneider2017earth}.
To demonstrate the use of high-fidelity simulations to improve ocean physics with AI, we will focus on two subgrid parameterization problems---ocean surface boundary layer (OSBL) vertical mixing and ocean mesoscale lateral mixing---specifically in the global configuration of the NOAA Geophysical Fluid Dynamics Laboratory (GFDL) ocean model. 

\section*{Learning ocean parameterizations for operational ocean models}

In this section, we will describe the datasets used to improve the parametrizations of ocean mixing, and the global operational models that we are targeting for implementation.  
We will also describe the range of ML models employed and their suitability for implementation in complex operational ocean models. 

\subsection*{Turbulent mixing parameterizations for the state-of-the-art GFDL OM4}

MOM6 (Modular Ocean Model version 6; \cite{adcroft2019gfdl}) is a state-of-the-art numerical ocean model developed primarily by the NOAA GFDL. 
It is designed to simulate ocean circulation and dynamics on a wide range of spatial and temporal scales. 
MOM6 is widely used in climate modeling, operational oceanography, and ocean process studies (e.g., \cite{harrison2022improved, wang2024improving}), and is used as the ocean component of the GFDL and NCAR climate models \cite{held2019structure}. 

Here, we employ GFDL OM4 (Ocean Model version 4), a global ocean and sea-ice configuration built upon MOM6, its oceanic core, and incorporating SIS2 (Sea Ice Simulator version 2) for sea ice dynamics and thermodynamics. 
OM4 has a nominal 1/4$^\circ$ horizontal resolution with 75 vertical layers to partially capture ocean mesoscale eddies \cite{adcroft2019gfdl}. 
This resolution is often referred to as eddy-permitting and resolves some mesoscale features, particularly in low latitudes. 
Yet, it is a much coarser resolution than is needed to even partially capture vertical boundary layer processes.  

\textbf{Ocean Vertical Mixing:} Ocean surface boundary layer (OSBL) vertical mixing impacts many aspects of the Earth's climate and weather. These impacts encompass both physical phenomena, such as tropical cyclones, El Niño--Southern Oscillation (ENSO), seasonal cycles, and long-term climate variability, as well as biogeochemical phenomena, including nutrient cycling and marine ecosystems (e.g. \cite{large_oceanic_1994,fox-kemper_ocean_2022}). 
Given the coarse vertical and horizontal resolution of the model, no vertical boundary layer processes are explicitly represented. 
There is a clean scale separation between resolved and unresolved vertical mixing physical processes; therefore, subgrid bulk parameterizations are often a reasonable approach. 
However, parameters associated with the subgrid closure are often ad hoc or weakly constrained by data.  
While there are successful examples of calibrating parameterizations with data-driven methods \cite{christopoulos2024}, this assumes that the original parameterizations is accurate. 

A common approach to studying the OSBL is to conduct Large Eddy Simulations (LES) that resolve vertical turbulent fluxes. LES generate high-fidelity data of ocean turbulence \cite{watkinsOSBL} and provide benchmarks for understanding physical processes in the OSBL. 
LES have also been used to develop and validate numerous OSBL parameterizations \cite{wagner2025formulation}, including several ML-based approaches \cite{ramadhan2023}.
In parallel, second-moment closure (SMC) is a sophisticated parameterization approach for OSBL turbulence that relies on Reynolds-averaging approaches to predict OSBL turbulent fluxes.
SMC reduces the representation of the OSBL from three spatial dimensions ($ x$, $ y$, $ z$) to one ($z$) by relying on empirical closure relationships between the turbulent fluxes (second moments) and mean flow properties.
SMCs predict vertical turbulent flux profiles that capture many key features simulated by the LES over multiple regimes \cite{umlauf2005second}. 

The bulk OSBL parameterizations commonly used in climate models rely on simplified closure assumptions compared to SMC.  
Instead of solving Reynolds-averaged equations for turbulent flux profiles, bulk parameterizations rely on empirical vertical shape functions to infer the turbulent flux profiles \cite{Reichl2018,large_oceanic_1994}.  
The shape functions for the diffusivity profiles in the existing bulk parameterizations are fixed in space and time, which is not supported by data. 

OM4 utilizes an energetically consistent Planetary Boundary Layer (ePBL) scheme \cite{Reichl2018}, which employs an energetic criterion to determine the mixing coefficient and diagnose the boundary layer depth.  
We set out to utilize machine learning to identify functional relationships that more accurately predict the vertical structure of vertical mixing for bulk OSBL parameterizations. 
By focusing on the ad hoc part of the bulk parameterization, we can improve the existing schemes without violating their underlying physical constraints (e.g., energetics-constrained vertical mixing, as in ePBL).

Both LES and SMC can be used to provide data for training machine learning models. SMCs are several orders of magnitude less computationally demanding than LES, allowing for significantly more OSBL training data to be generated and to span the parameter regime more widely and densely. 
Here, we primarily used SMC data, validated against LES, to train our AI-based parameterizations of ocean vertical mixing \cite{sane2023}. 

\textbf{Ocean Lateral Mixing Parameterizations:} Mesoscale eddies are crucial in determining large-scale ocean transport, particularly by exchanging energy and momentum with the mean flow. 
Lateral mixing in the ocean is driven by mesoscale eddies that emerge on the spatial scale of the first baroclinic Rossby deformation radius (10--100 $\mathrm{km}$) \cite{hallberg2013using}. The size of these eddies roughly corresponds to the horizontal grid spacing of existing global ocean models.  
Thus, unlike for ocean vertical mixing, there is no clear scale separation between parameterized and resolved processes. 
In this regime, often referred to as the gray zone, traditional physics-based parameterizations have low accuracy in capturing the subgrid fluxes' impact on the large-scale. 
For example, in OM4, downgradient momentum fluxes are traditionally used to parameterize the direct cascade of energy or enstrophy at the grid-scale \cite{griffies2000biharmonic, bachman2017scale}. Several studies have proposed upgradient momentum fluxes to return the spuriously dissipated energy (backscatter; \cite{chang2023remote}). 
However, both approaches rely on the assumption that there is a statistical correlation between the gradient of the mean flow and subfilter fluxes, which is not backed by data \cite{nadiga2008orientation}.
When scale separation does not hold, we can directly learn the relationship between the resolved flow and subfilter fluxes from the data, without relying on the flux-gradient relation. 

For the development of mesoscale eddy parameterizations, we have relied on a range of simulations that resolve at least the first deformation radius.
This includes simple quasigeostrophic models, which primarily capture the interactions between ocean mesoscale eddies and the large-scale flow in idealized settings \cite{ross2022benchmarking}, primitive equation models in idealized geometric configurations \cite{Bolton2019, zanna2020data}, and state-of-the-art coupled climate models which include a host of other interactions, including with the atmosphere and sea ice \cite{griffies2015impacts, guillaumin2021stochastic}. 
Unlike for SMC data, here we need to filter and coarse-grain the data to diagnose the subgrid forcing that a coarse-resolution model at a given resolution would be lacking, and that needs to be parameterized \cite{guan2022stable} (see Figure~\ref{fig:subgrid}). 
In the next section, we describe three data-driven parameterizations of momentum fluxes generated by mesoscale eddies, which differ in their computational complexity and the amount of imposed physical constraints.

\subsection*{Spectrum of ML machinery for parameterizations} 

\begin{figure}
    \centering
    \includegraphics[width=1\linewidth]{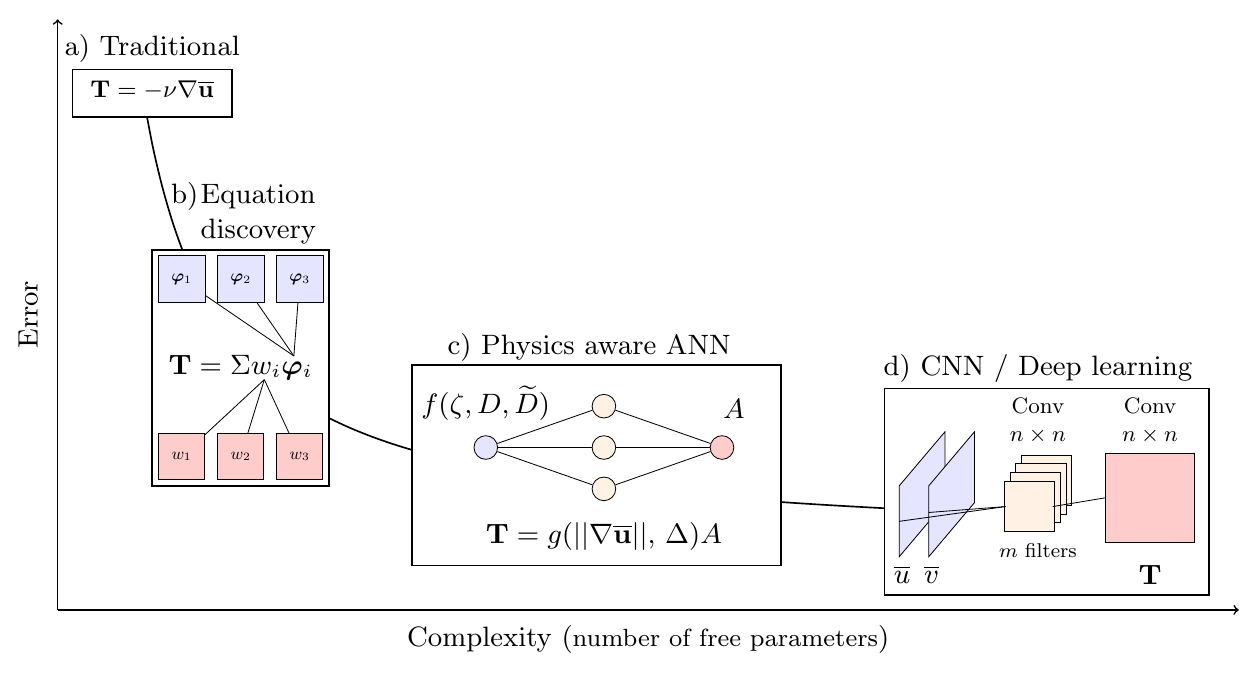}
    \caption{Schematic of Structural Error vs. Complexity. Panel a--d are organized on a structural error vs complexity curve, which generally transitions from high error and low complexity (e.g., small number of free parameters) to low error and high complexity. Here, each panel uses the case of parameterizing the momentum stress tensor ($\mathbf{T}$) in different ways as an example. Panel a) illustrates a more traditional approach to flux parameterization, where the stress tensor is simply a function of the velocity gradient, for anti-viscosity \cite{chang2023remote}. Panel b) shows equation discovery methods, where the basis functions $\bm{\varphi}$ may be chosen or discovered functions of the velocity field, such as vorticity $\bm{\varphi}_1 = \bm{\zeta} = \overline{\bm{\nabla}}\times\overline{\mathbf{u}}$, shearing deformation $\bm{\varphi}_2=\mathbf{D} = \partial_y\overline{u} + \partial_x\overline{v}$, and the stretching deformation $\bm{\varphi}_3 = \widetilde{\bm{D}} = \partial_x\overline{u} - \partial_y\overline{v}$ \cite{zanna2020data,ross2022benchmarking}. Panel c) shows physics-aware ANNs, where as an example the inputs are also hand chosen (e.g., $\bm{\zeta}$, $\mathbf{D}$,  $\widetilde{\bm{D}}$), and the outputs from the network are dimensionalized using some physics-informed function $g$ of an appropriate norm of the input fields, $||\bm{\nabla}\overline{\mathbf{u}}||$ and the grid-spacing $\Delta$. Panel d) shows CNNs where the input fields are chosen to be some variables, e.g., filtered velocity fields, with minimal inductive bias. }
    \label{fig:complexity}
\end{figure}

Generally, most physics-based parameterizations capture a simple bulk relationship between the effect of subgrid forcing and the large scale. 
The relationship is typically encapsulated in a convenient mathematical form that is interpretable. 
ML models, in particular, deep learning methods, often lack this inherent interpretability. 
Here, we have utilized ML models of varying complexity and interpretability, as illustrated in Figure~\ref{fig:complexity}, to investigate the capability of data-driven methods to be easily incorporated in operational climate models and improve the prediction of subgrid ocean parameterizations. 


On the most sophisticated and computationally expensive end of the ML spectrum, deep learning methods, such as convolutional neural networks (CNNs), are capable of capturing the complex relationships between the subgrid fluxes and resolved fields. 
These models are particularly useful when it is not a priori clear what the minimal set of input variables or spatio-temporal nonlocality is required, and there is no compelling physical reason to impose an inductive bias by hand-designing input features or choosing a problem-specific model architecture. 
It has been shown that CNNs perform well, with or without physical constraints, in learning the relationship between the subgrid ocean momentum fluxes and resolved flow in both idealized ocean configurations \cite{Bolton2019,zanna2020data,ross2022benchmarking, yan2024choice} and realistic ocean settings \cite{guillaumin2021stochastic, gultekin2024analysis}. 
In addition, CNNs implemented as a parameterization in an idealized ocean model can energize the flow appropriately \cite{zhang2023implementation} (see also Figure~\ref{fig:gz21_global}). 

However, several challenges are associated with CNNs that require careful consideration. Conventional CNNs are composed of millions of parameters and require a large number of matrix operations at inference time, which can impose a computational and memory bottleneck when coupled to a forward numerical model that runs on CPUs. Additionally, efficient implementations for CNN inference require sophisticated and flexible code that already exists in Python libraries (such as PyTorch and TensorFlow), and porting this to Fortran can increase the software development and maintenance overhead. Zhang et al. addressed these using FORPY \cite{zhang2023implementation} and also experimented with SmartSim (HPE; \cite{parteeetal}), which provides an interface for leveraging GPUs and calling Python functions from Fortran. 
Another set of issues arises from the lack of awareness about the computational grid and physical boundaries in conventional CNN designs. It has been observed that CNNs often attempt to learn relationships akin to spatial gradients \cite{zanna2020data}. However, without providing grid metrics, there is no guarantee that the learned spatial operations will scale appropriately when the grid resolution changes or becomes non-uniform. Additionally, grid-unaware CNNs cannot be expected to learn or be designed to be aware of physical symmetries and invariances \cite{pawar2020priori}. Consequently, these architectures cannot distinguish arbitrary boundaries and impose physically appropriate boundary conditions. 
\cite{zhang2024addressingoutofsampleissuesmultilayer} has shown that ad hoc approaches can help partially alleviate issues with boundaries, and \cite{zanna2020data, guan2022learning} showed that some physical constraints can be built into CNN layers. 


In principle, CNNs could learn the connections between outputs and appropriate aspects from the input data; for example, aspects like dependence on spatial derivatives \cite{zanna2020data} or requirement of limited spatial non-locality \cite{gultekin2024analysis} can be inferred when the input data is chosen to be a superset of all these aspects. Thus, in scenarios where the parameterization developer has a good idea of which specific aspects to include, based on physical knowledge or lessons learned from probing CNN architectures, utilizing simple artificial neural networks (ANNs), also known as multilayer perceptrons (MLPs), can be beneficial. ANNs have lower design complexity than CNNs, and with limited inputs, can be computationally more efficient. Also, their relatively simple architecture often makes it relatively straightforward to incorporate numerous potentially useful design choices, such as physical symmetries and invariances, non-dimensionalization approaches, grid awareness, limited and controlled spatial non-locality, and physically justified boundary condition choices \cite{perezhogin2025generalizable, balwada2025design}. 
ANNs have low software complexity, and their inference modules can be easily implemented in Fortran or any other language and optimized to have minimal computational overhead, even on CPUs. 
Given their simplicity, these architectures have been a popular choice in several ML-based parameterizations, including atmospheric parameterizations of vertical processes \cite{yuval2020stable,rasp2018deep}, oceanic boundary layers \cite{sane2023}, and lateral oceanic processes \cite{perezhogin2025generalizable}. In a sense, ANNs are more interpretable than CNNs, as very specific choices have been made in selecting inputs; however, the exact non-linear relationships learned are implicit and not easily decipherable.

Ultimately, a more interpretable approach can be realized by discovering equations for the subgrid parameterizations. 
Rather than redefining the functional form of the parameters or parameterizations, we learn it from data as an equation.  
For lateral momentum fluxes, Zanna \& Bolton \cite{zanna2020data} and Ross et al. \cite{ross2022benchmarking} discovered an expression that depends on the strain and vorticity of the resolved flow, which can be scaled with resolution \cite{ross2022benchmarking,perezhogin2025generalizable}, using Bayesian sparse regression and genetic programming algorithms. 
For the vertical mixing profile, Sane et al. (2025) \cite{Sane2025} found relationships for the vertical profile of diffusivity (structure) and its magnitude using equation discovery and empirical fitting, distilling the results from an ANN. The vertical structure and amplitude of diffusivity align with SMC and are therefore constrained by more accurate physics than those employed in first-order closures used in ocean climate models. Recently, Grundner et al. (2025) \cite{grundner2025} distilled a functional form of a cloud cover parameterization from a neural network. 

\begin{figure}
    \centering
    \includegraphics[width=1\linewidth]{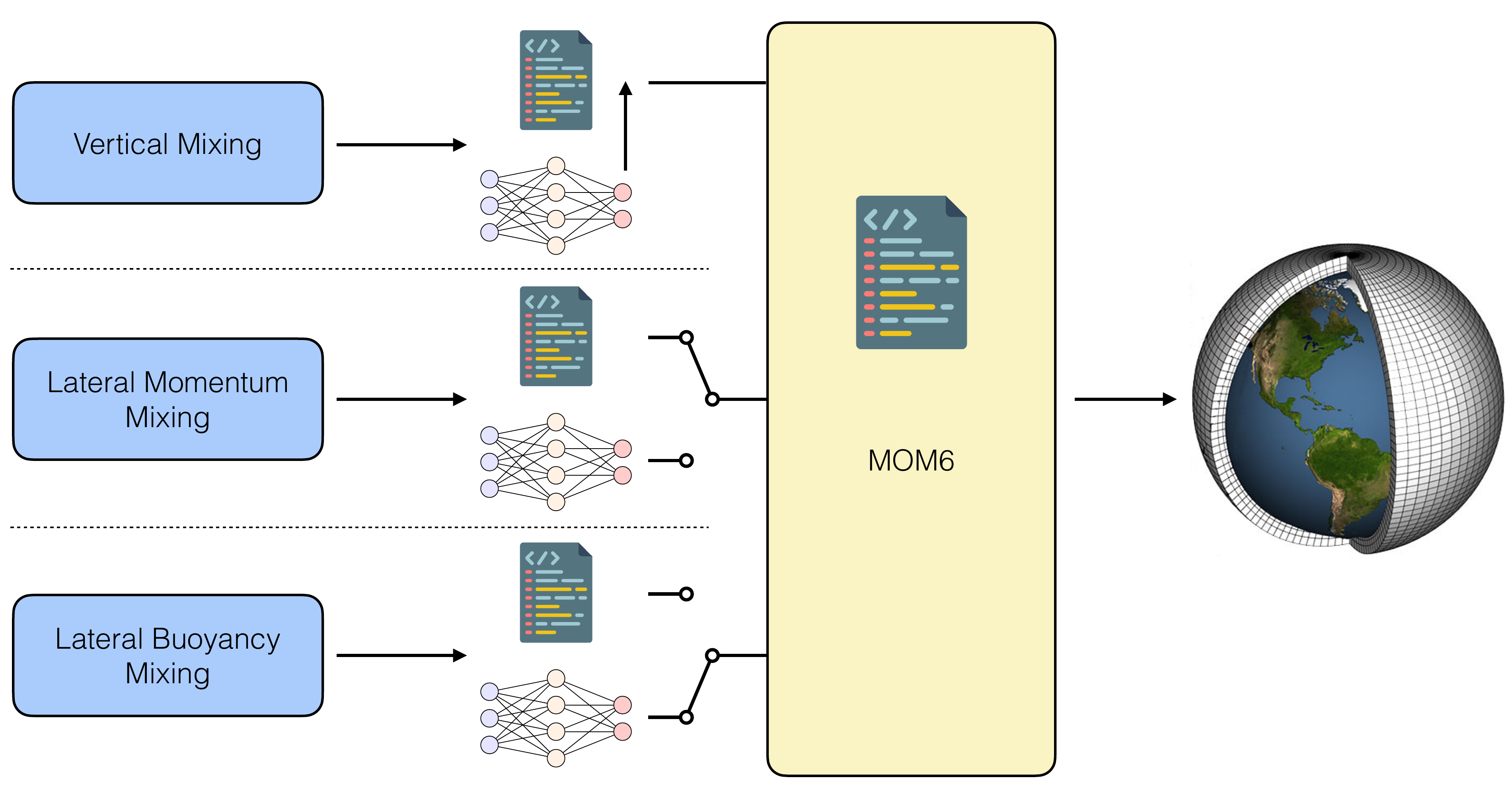}
    \caption{Schematic of an ANN module, written in Fortran, in MOM6 being used for multiple parameterizations within the GCM.  The ANN module is created to perform the inference for multi-layer ANN models trained in PyTorch or other ML libraries. The weights, biases, and others parameters are stored in a NetCDF file, which communicate the information to the Fortran module. For ocean vertical mixing, the module computes the vertical profile of diffusivity, while for the lateral momentum mixing parametrization it computes the full eddy momentum flux. 
    }
    \label{fig:ANN_module}
\end{figure}

The ANNs and equations for ocean turbulence closures have been implemented directly in Fortran (the language of the existing climate model).  
Specifically, we created a module in  MOM6, written in Fortran, which performs inference for multi-layer ANN models trained in PyTorch or other ML libraries (see schematic in Figure \ref{fig:ANN_module}). The weights and biases, as well as auxiliary information (normalization parameters), are stored in the NetCDF file, which passes the information to the Fortran module. 
While it is not necessary for the ANNs and the equations to be written as a Fortran module, it has several advantages when working with existing climate simulations, which are written in Fortran. 

The key component of the ANN inference---matrix multiplication---is designed to be reproducible. 
Reproducible computations ensure that the result is always identical at a bitwise level, regardless of the compilation options. 
The reproducibility of computations is an essential feature of the MOM6 code, which is heavily utilized in its testing system. 
The ANN Fortran module enables inference out of the box, eliminating the need for installing dependencies or GPUs. 
The implemented ANN parameterizations of vertical mixing and mesoscale eddies utilize this Fortran module.
Note that the parametrizations can be seamlessly integrated into any simulation, whether CPU- or GPU-based, using the weights.

\section*{AI-enhanced ocean climate simulations}

We have successfully implemented multiple data-driven parametrizations in OM4, using the framework described above. 
In this section, we briefly examine the impact of data-driven parameterizations on the mean biases in forced OM4 simulations using JRA forcing for vertical mixing parameterizations and CORE-II AIF forcing for lateral mesoscale momentum parameterizations.  
For both vertical and lateral mixing parameterizations, we employ an ANN and equation-based ML models, learned from data. 
In ocean-forced and coupled climate simulations, numerous biases in sea surface temperatures (SSTs) and mixed layer depths (MLDs) are shared. 
Some of the culprits for these biases are the subgrid parameterizations used in ocean models, particularly those related to vertical mixing in the OSBL or mesoscale eddy fluxes. These biases can have a significant impact on the large-scale circulation and its response to external forcing. 

We would need long coupled atmosphere-ocean-sea-ice runs to assess the full impact of the data-driven closures on the ocean circulation, which is beyond the scope of this paper, as we aim to present solely a framework for implementing data-driven parameterizations in coupled ocean-sea-ice operational models. 
Therefore, we focus solely on the near-surface properties — primarily SSTs and MLDs — to illustrate the impact of the data-driven parameterizations on model solutions. In general, the effect of parameterizations on these fields is already visible after several years of integrating forced-ocean models, such as OM4. Here we run the simulations over 60 years, and analyze the last 15 years of the simulations.
We concentrate on two data-driven parameterizations previously presented: vertical mixing, utilizing AI-enhanced diffusivity as part of the ePBL scheme, and ocean lateral momentum fluxes, employing an ANN and ZB20 \cite{zanna2020data}.  

\begin{figure} 
	\centering
      \includegraphics[width=0.7\textwidth]{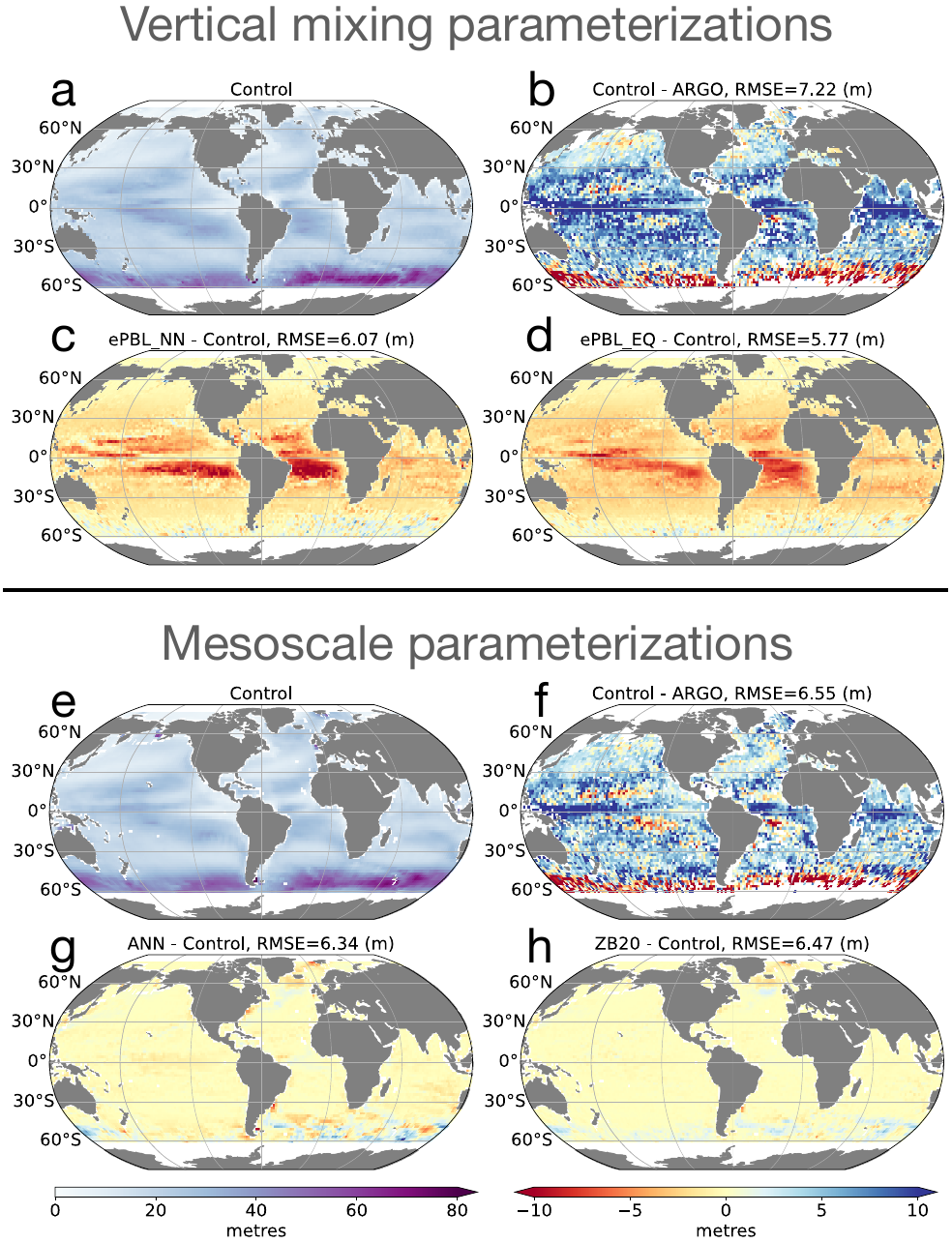}
	\caption{\textbf{Summer Mixed Layer Depth (MLD) in meters: Changes induced by ML parameterizations for ePBL and lateral momentum fluxes.} (\textbf{A, E}) Summer MLD, found using the potential anomaly criterion \cite{ReichlMLD}, in the control run for ePBL with JRA forcing and for mesoscale parameterization with CORE IAF, respectively; (\textbf{B}) Bias of control with respect to ARGO; (\textbf{C}) Difference between ePBL\_NN and control; (\textbf{D}) Difference between ePBL\_EQ and control; (\textbf{F}) Difference between ANN and control; (\textbf{G}) Difference between ZB20 and control.}
	\label{fig:mld-summer} 
\end{figure}

Using the baseline ePBL scheme implemented in OM4, the summer MLDs are deeper in the Tropics than those observed in ARGO, see Figure \ref{fig:mld-summer}(b). Using results from Sane et al. (2023, 2025) \cite{sane2023} \cite{Sane2025},  either neural networks or equations learned from data can be used to replace the diffusivity in the baseline ePBL scheme. The data-driven diffusivities lead to a significant reduction in summer MLD biases, particularly in the Tropics and Subtropics (Figure \ref{fig:mld-summer}c,d). 

\begin{figure}
	\centering
        \includegraphics[width=0.7\textwidth]{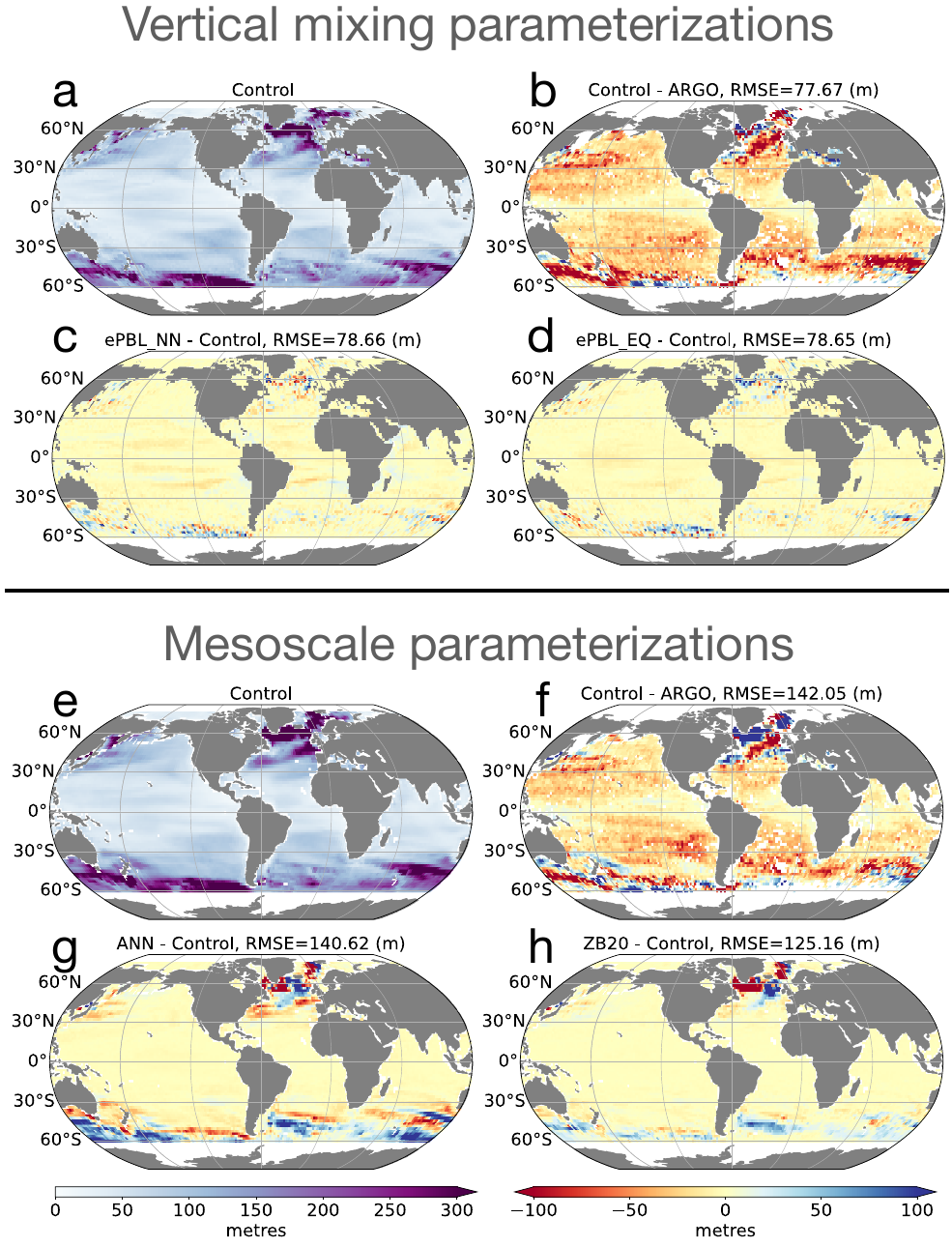}
	\caption{\textbf{Winter Mixed Layer Depth: Changes induced by ML parameterizations for ePBL and lateral momentum fluxes.} Panels are the same as Fig. \ref{fig:mld-summer} but for winter MLD. Note that control biases exhibit the same patterns under different forcing conditions (JRA vs. CORE IAF, but with varying magnitudes.}
	\label{fig:mld-winter} 
\end{figure}

The AI-enhanced diffusivity only replaces the ad hoc components of the baseline ePBL scheme, maintaining the original physics constraints in the original baseline ePBL parameterization \cite{Reichl2018}. 
Due to the inexpensive generation of training data, the neural networks and equations capture more comprehensive interactions between mixing and forcing, than the profile used in standard parameterizations (e.g., ePBL or KPP). 
The resulting reduction in biases is due to deficiencies in the baseline's ePBL diffusivity, which lacked the appropriate sensitivity to surface heating. 
Despite the above enhancements, the AI-enhanced diffusivity only improved the tropical mixed layer depths, and these improvements were only observed during the summer, with no noticeable changes in the winter \ref{fig:mld-winter}c,d.  

Mesoscale lateral momentum parameterizations, often referred to as backscatter schemes, reinject energy into larger scales, enabling more efficient extraction of available potential energy via resolved eddies and thereby modifying ocean stratification.
Both ANN (Perezhogin et al. 2025 \cite{perezhogin2025generalizable}) and equation-based parameterizations (ZB20, Perezhogin et al. 2024 \cite{perezhogin2024stable}) are implemented in OM4. In contrast to the AI-enhanced ePBL, we are not simply modifying a parameter but rather adding a tendency to the model equation, which represents the effect of the mesoscale eddy flux divergence on the resolved state. 
The response in summer MLD for the backscatter parameterizations is relatively muted, compared to the response when introducing the data-driven ePBL parameterization (Figure \ref{fig:mld-summer}g,h). We identify a small reduction in biases in high latitudes in the North Atlantic and Southern Ocean. This muted response is somewhat expected as eddy momentum fluxes are not strong in the summer or in the Tropics. 

The response in winter MLD is more significant for the momentum parameterizations at mid- and high-latitudes, where eddy activity is more pronounced. The pattern of response is correlated for both parameterized models (Figure \ref{fig:mld-winter}g,h), with seemingly a stronger impact from the ANN parametrization compared to ZB20, except in the North Atlantic. 
Both parametrizations reduce the biases in MLD in the North Atlantic and in parts of the Southern Ocean (particularly south of Australia); however, some biases are reinforced in the South Atlantic. 
The ZB20 parameterization exhibits the best global bias reduction, particularly due to its effect in the North Atlantic.

\begin{figure} 
	\centering
  \includegraphics[width=0.7\textwidth]{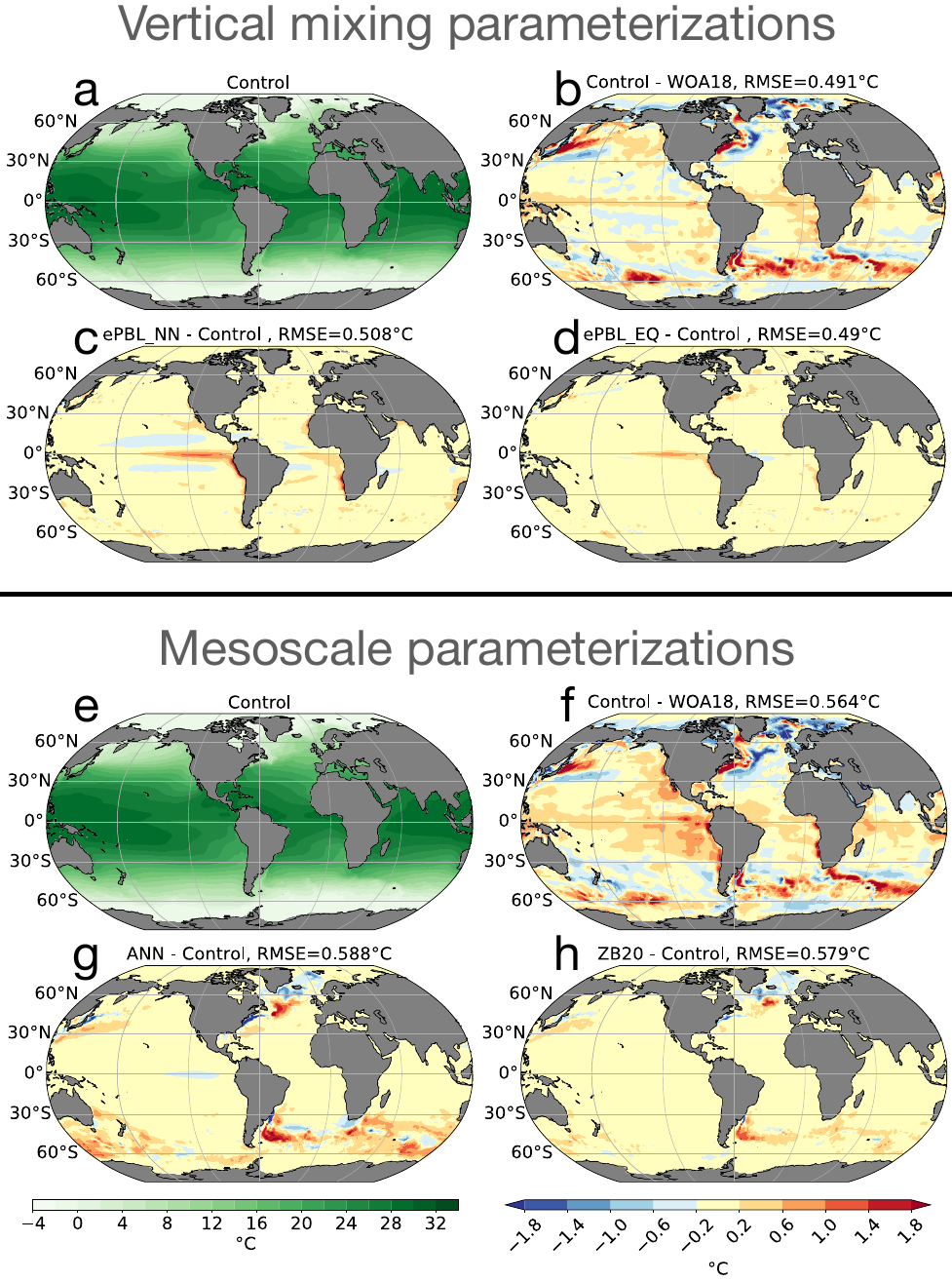}
	\caption{\textbf{Annual Mean Sea Surface Temperatures in $^{\circ} C$.} Same captions as Fig. \ref{fig:mld-summer}. Note again that the biases in SSTs under different forcings are shared by the models, except in the Tropics. The repeat forcing from IAF lead to larger biases in the ENSO region in particular, as expected}
	\label{fig:ePBL_SST} 
\end{figure}

The data-driven parametrizations also impact SSTs. In the tropics, the AI-ePBL parameterizations (ANN and equation discovery) appear to have increased the bias in annual mean SSTs (Figure \ref{fig:ePBL_SST}c-d), despite reducing biases in MLD and stratification. However, the global mean biases remain unchanged or are only marginally degraded. However, the SST response is pronounced when the lateral momentum flux parameterizations are applied.  

The mesoscale eddy parameterizations influence the biases in mid- and high-latitudes (Figure \ref{fig:ePBL_SST}g-h), where lateral eddy fluxes impact the large scale. The strongest response is obtained when the ANN parameterization is employed, rather than the equation discovery (the latter being sparse and therefore less expressive compared to the ANN). 
The ANN parameterization reduced the biases in the North Atlantic and North Pacific, primarily by affecting the separation and strength of the Gulf Stream and Kuroshio boundary currents, respectively. For example, this change lead to more accurate northward heat transport, and a reduction of the warm bias along the Gulf Stream and the cold bias in the North Atlantic. However, the ANN parameterization further increases the warm bias in the Southern Ocean, similarly to traditional anti-viscosity lateral momentum parameterizations \cite{juricke2020ocean, chang2023remote}. 
The ZB20 parameterization, however, reduces the biases in the North Atlantic (albeit with a slightly weaker amplitude compared to the ANN), but does not deteriorate the Southern Ocean  Figure \ref{fig:ePBL_SST}g-h). 

We have demonstrated that AI methods can effectively learn subgrid physics for a given process using high-resolution simulations, and shown that AI-based subgrid parameterizations can be integrated into state-of-the-art ocean models. 
Yet, we still face two hurdles: 1) many physical processes are missing, and this means that we may not always be improving simulations, and 2) we are not directly learning information from observational data but rather from potentially imperfect simulations.

\subsection*{Learning data assimilation increments for bias correction and parameterizations}

Learning subgrid parameterizations from high-resolution simulations has many advantages as described above.
Yet, these methods will only target specific processes for which we have enough high-quality data. 
Other sources of model bias, including missing or poorly represented physical processes, poorly calibrated model physics parameters, and numerical errors, still require attention. 
A significant body of work from the Data Assimilation (DA) community has shown that the corrections, or increments, applied to a numerical simulation during DA are essentially a manifestation of these various sources of model bias and may be predictable \cite{Rodwell2007,Palmer2011}. Subsequently, ML provides a unique opportunity for deriving state-dependent representations of DA increments \cite{Gregory2023,Verma2025}, including its simplest form, known as nudging \cite{Chapman2025}. 
Through this framework, the ML model serves as an emulator for the numerical model's systematic errors, allowing it to provide a state-dependent bias correction for subsequent simulations. This hybrid ML-based bias correction approach has shown promise in idealized models \cite{Brajard2020,Brajard2021,Farchi2021,Mojgani2022}, and is now being scaled up to global atmosphere and ocean-sea-ice simulations \cite{Bonavita2020,Watt2021,Chen2022,Chapman2025,Gregory2024,Gregory2025}.
\begin{figure} 
	\centering
\includegraphics[width=0.9\textwidth]{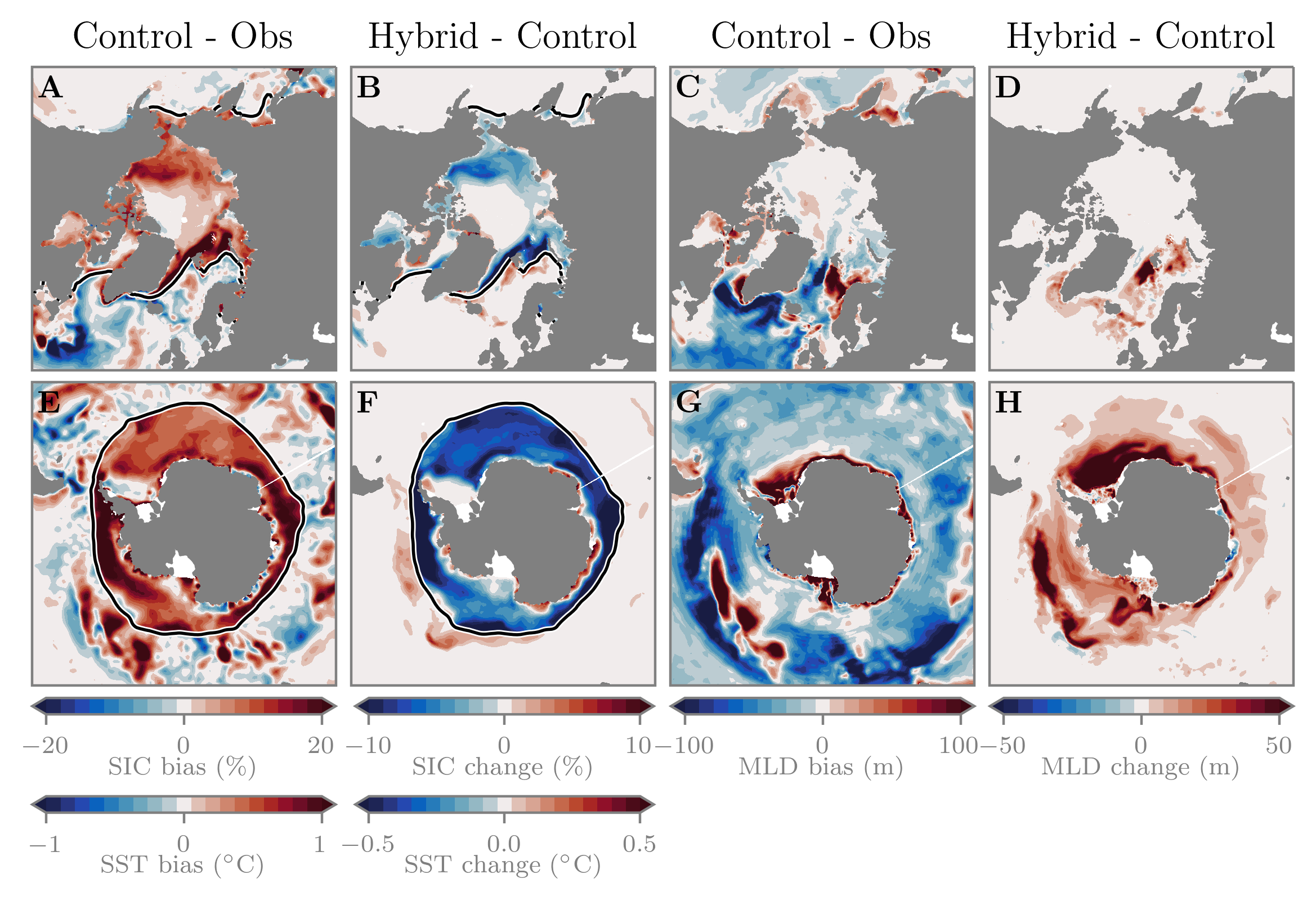} 
	\caption{\textbf{Changes in online ocean and sea ice biases from machine-learned sea ice bias correction.}
		(\textbf{A}) Arctic sea ice concentration (SIC) and sea-surface temperature (SST) biases from a 5-year control simulation (2018--2022). The model's climatological sea ice edge is shown by the black contour. Poleward of this contour are SIC biases, and equatorward are SST biases. (\textbf{B}) The change in SIC and SST due to ML-based sea ice bias correction. Note that the colorbar is half the magnitude of Panel \textbf{A}. (\textbf{C}) Mixed-layer depth (MLD) biases in the control simulation. (\textbf{D}) The change in MLD due to ML-based sea ice bias correction (positive values indicate a deeper mixed layer in the hybrid simulation). (\textbf{E}--\textbf{H}) Same as \textbf{A}--\textbf{D} but for Antarctic.}
	\label{fig:Seaice} 
\end{figure}

Figure \ref{fig:Seaice} highlights a case where ML was used to online bias-correct the sea ice concentration (SIC) errors in a 5-year global ice-ocean simulation \cite{Gregory2024}. 
The CNN-based ML model has approximately $10^5$ weights and takes spatially local inputs from the sea ice, ocean, and atmosphere states on a 9$\times$9 stencil, outputting a prediction of the model error in SIC at each grid point. 
It is therefore taking advantage of the non-local information coming from the interaction of different variables, and can serve as a measure of memory. The CNN was coded directly into the Fortran source code of the GFDL sea ice model SIS2; hence, the 9$\times$9 stencil size was intentionally chosen to match the default halo size used in parallelized GFDL OM4 simulations. With this configuration, the computational performance is comparable to that of the free-running numerical model, incurring a 0.3\% slowdown due to ML inference on the CPU (we discuss the computational impact of large ML models in the next section).
In Figure \ref{fig:Seaice}, SST and SIC biases are relative to version 2.1 of the NOAA Optimum Interpolation (OISST) product \cite{Huang2021}, and the National Snow and Ice Data Center NASA Team passive microwave data set \cite{DiGirolamo2022}, respectively. 
The MLD bias shown in Figures \ref{fig:Seaice}C and \ref{fig:Seaice}G  corresponds to the difference between the free-running model's 5-year climatology MLD and a 52-year climatology from observations (here a climatology from ARGO and World Ocean Database observations collected over the period 1970--2021).
The free-running model generally exhibits positive sea ice biases in both hemispheres, with some regions displaying locally negative biases, such as the Laptev Sea in the Arctic and around the Antarctic coastline (Figures \ref{fig:Seaice}A,E). The ML model generalizes to both hemispheres and systematically reduces Arctic and Antarctic SIC biases---adding sea ice in regions of negative bias, and removing sea ice in regions of positive bias (Figures \ref{fig:Seaice}B,F). However, the hybrid approach's ability to improve ocean model biases yields mixed results. Changes in SST are muted because both simulations employ strong nudging of SSTs toward observations. Meanwhile, MLD biases are degraded in some regions (e.g. positive (deep) MLD biases are exacerbated in the Weddell Sea in the Antarctic and the Norwegian Sea in the Arctic) and improved in others (e.g. negative (shallow) MLD biases are improved through much of the Southern Ocean and the Greenland and Labrador Seas in the Arctic). These mixed results for ocean variables may be exposing compensating ice-ocean model errors and highlight the complexity of using ML schemes in coupled models. Learning both ice and ocean increments together in a strongly-coupled DA framework could be a potential pathway for addressing this issue. 

One additional takeaway from \cite{Gregory2024,Gregory2025} was the potential of this DA+ML framework to enhance the online generalization of ML schemes. It is often the case that an ML model that achieves high prediction skill on offline validation data performs poorly when implemented into a numerical climate model \cite{rasp2018deep,ross2022benchmarking}. 
This typically occurs because the ML model is continuously generating new climate states online that it has not encountered during training (i.e., out-of-sample problem). 
A commonly referenced solution to this problem is the process of ``online training," where the weights of the ML model are updated during integration of the numerical model. Several works \cite{um2020,Kochkov2024} have demonstrated the potential benefits of this strategy for the long-term accuracy and stability of ML methods. 
The downside of this approach is that it requires differentiable numerical model code, which is not the standard for any current IPCC-class climate models. 
While some studies have found success in rewriting climate model code in a differentiable programming language \cite{Kochkov2024}, this is impractical for modeling centers to adopt immediately.
The approach detailed in \cite{Gregory2024,Gregory2025} is a natural way to generate training data that captures the feedback between the ML model and the numerical climate model, without requiring differentiable code. 
This is achieved by running a new simulation where, at each assimilation time, the model state is first corrected using the offline-trained ML model. Then, DA is used to update this ML-corrected state. 
The DA increment, therefore, reflects the residual error from the ML model; thus, the sum of the DA and ML increments provides new information for the ML model to target with subsequent offline training. 
This approach can be performed iteratively until the DA increments converge on zero-mean Gaussian noise and the ML increments converge on the systematic model errors. 
Similar strategies of data augmentation can also be applied to data-driven AI parameterizations using high-resolution simulations.
An alternative approach to address the differentiability requirement is to approximate the gradients of the numerical model. 
This is achieved by training an approximate differentiable surrogate DL model (emulator), which can then provide gradient information for online learning without modifying the original numerical model code \cite{frezat2023gradient}. 
While this approach shows promising results and offers an interesting perspective for training ML models with existing climate model implementations, its effectiveness remains dependent on the quality of the emulator approximation, which may pose challenges when applied to more complex dynamical systems, as discussed in the next section.

\subsection*{Current Challenges and Future Directions} 

We have demonstrated a framework for learning from data, either by utilizing high-resolution simulations to learn a subgrid process missing from coarse-resolution simulations, or by leveraging data-assimilation increments that capture a multitude of processes within a given coarse-resolution model relative to observations. 
While the first method captures the direct effect of physics, and the second targets model bias, we need to continue exploring both approaches simultaneously, namely, by improving the model physics to reduce model biases.  
We must therefore pursue several avenues to address these goals, which include: 1) using larger AI models; 2) building generalizable AI models; 3) calibration or tuning of climate models. 
For each approach, we will also discuss the challenges associated with the methods. 

\subsubsection*{Large ML models in realistic global models}

To enhance the fidelity of global simulations using AI parameterizations, one strategy is to increase model complexity and capability, often resulting in larger ML models.
However, this comes with higher computational costs.
For instance, implementing the stochastic deep learning model from Guillaumin and Zanna, 2021 \cite{guillaumin2021stochastic}, hereafter GZ21,
for mesoscale lateral momentum mixing in MOM6, trained on CM2.6 surface velocity data, improved solutions in an idealized double gyre setup \cite{zhang2023implementation}.
A global test using the OM4 configuration \cite{adcroft2019gfdl} remained stable for a year, though inference costs limited longer integrations.
Nevertheless, the global test emphasized two key challenges: nonlocal effects and inference costs.

Lateral mixing eddy parameterizations must account for nonlocal dynamics, especially near coastal boundaries.
CNNs can capture such effects if their receptive fields are sufficiently large. GZ21 used a $21 \times 21$ stencil and showed strong offline and online performance. However, Gultekin et al. \cite{gultekin2024analysis} found that a $7 \times 7$ stencil was sufficient for offline tests, although online validation is still required.

A major nonlocal challenge is handling coastal interactions.
CNNs struggle with the vast number of possible land/ocean configurations (here, $2^{N^2}$ for an $N \times N$ stencil) leading to extrapolation errors when unseen permutations arise.
Zhang et al., 2023 \cite{zhang2023implementation}, observed boundary artifacts when GZ21 was trained on open-ocean data and applied near coasts.
Zhang et al. (2025) \cite{zhang2024addressingoutofsampleissuesmultilayer} mitigated this issue using zero and replicate padding; however, a more robust solution involves training on global datasets that include coastal dynamics.
This requires high-quality, multiscale data and network architectures capable of learning from it without overfitting or underfitting, which is not readily available. 

Even modest CNNs, such as GZ21, are computationally expensive.
In idealized MOM6 tests, CNN inference on CPUs was an order of magnitude slower than the dynamical core, while GPU inference reduced this to 10\% of the core's cost.
On Frontier (38 nodes, 152 GPUs), inference time was ~6\% of the core's, though GPU utilization was low due to the small model size.
Despite this, GPU-based inference is essential for practical performance.

To improve efficiency, dedicated inference nodes could be used in CPU-dominated simulations, trading some latency for better hardware utilization.
Partee et al. \cite{parteeetal} demonstrated that with modern interconnects, this overhead is minimal ($\sim$1–2\%).

For GPU-limited platforms, alternatives like knowledge distillation are promising.
Here, a smaller “student” model, possibly even a closed-form equation (e.g.Sane et al., 2025 \cite{Sane2025}), is trained to emulate a larger “teacher” model.
This approach is particularly relevant for CPU-based climate and weather systems and aligns with the broader goal of striking a balance between nonlocal fidelity, interpretability and computational efficiency.

\subsubsection*{Generalization across climates, regimes, and resolutions}

A central challenge in the development of ML parameterizations lies in achieving generalization across varying spatial resolutions, dynamical regimes, and external forcings. 
While physical models are constructed in most general terms and through considering extremes or edge cases, ML models are not constructed to hold across a broad range of conditions but to perform optimally on specific datasets. 
Consequently, ML models can produce inconsistent or unphysical predictions when tested in scenarios substantially different from those on which they were trained, including warmer climates or unseen circulation patterns. 
The scarcity of observational or high-fidelity data representative of a changing climate limits the robustness of na\"ive ML approaches to climate prediction despite substantial power to infer from data patterns corresponding to physical processes \cite{yuval2021stable}. 

The multi-scale nature of climate also motivates parameterizations adaptable to different computational grids.
However, changes in spatial resolution can alter the structure of inputs to ML parameterizations, undermining model accuracy \cite{yuval2023neural}. Further, resolving physical scales outside those of training data may invalidate learned feedbacks, leading to unstable or unrealistic behavior \cite{eyring_pushing_2024, perezhogin2024stable}.
Additionally, data-driven ML models integrated into physical models frequently suffer from grid mismatches. Simple interpolation or averaging of physical quantities onto the ML input space can reduce accuracy and induce numerical instabilities \cite{zhang2024addressingoutofsampleissuesmultilayer}. These generalization failures can manifest as sudden instabilities (blow-ups) in numerical simulations or as more subtle effects that are challenging to diagnose.
Several strategies have been proposed to mitigate these issues.

As a solution to generalization, the community has long discussed physical constraints, including conservation laws, similarity relationships, and symmetry.
These may be introduced as soft constraints by penalizing violations of the constraint during optimization \cite{frezat2021physical}.
Alternatively, these may be incorporated as hard constraints on model architecture, e.g., to guarantee translational or scale invariance \cite{zanna2020data,wang2021incorporatingsymmetrydeepdynamics}.
However, physical constraints are not sufficient to ensure generalization \cite{barsinai2019discretizations}, and in some cases, they can even degrade performance or stability \cite{frame_strictly_2023}. 


Other, and perhaps more promising, avenues involve modifying the learned relationship through physical insight.
Simply reformulating model outputs as fluxes instead of tendencies may ensure conservation of mass and energy \cite{yuval2021stable}, and therefore more appropriate for generalization.
Similarly, improvements follow other physics-informed input transformations, \emph{e.g.} recasting variables such as temperature and humidity into a more climate-invariant form, moist static energy, through trial and error,  has been shown to improve robustness under warming scenarios \cite{beucler2024climateinvariant}.
Finally, more formal dimensionless normalization of inputs and outputs, as in ref. \cite{perezhogin2025generalizable}, may afford generalization across a broader range of dynamical regimes and grid resolution. 

\subsubsection*{Toward Fully Coupled Simulations, and the role of autoregressive ML emulators for tuning}

Understanding how ML-based parameterizations impact long-term past, present, and future climate relies on coupling hybrid ocean and sea ice models to interactive atmosphere and land components. 
Our framework to learn subgrid physics or model corrections from data can be tested, to some extent, in fully coupled climate simulations. 
However, this presents a whole new set of challenges. 
In forced OM4 configurations, changes to the OSBL and mesoscale lateral momentum fluxes through machine-learned subgrid closures have no impact on the atmosphere, which likely helps maintain stability over long integrations, and this is similar for the ML sea-ice corrections in the SPEAR forced setup. 
In fully coupled simulations, these schemes have the ability to flux heat, moisture, and momentum to the atmosphere, which may trigger convective or cloud-based feedbacks, subsequently pushing the ML model out of sample and causing runaway behavior. 

\begin{figure}
	\centering
\includegraphics[width=\textwidth]{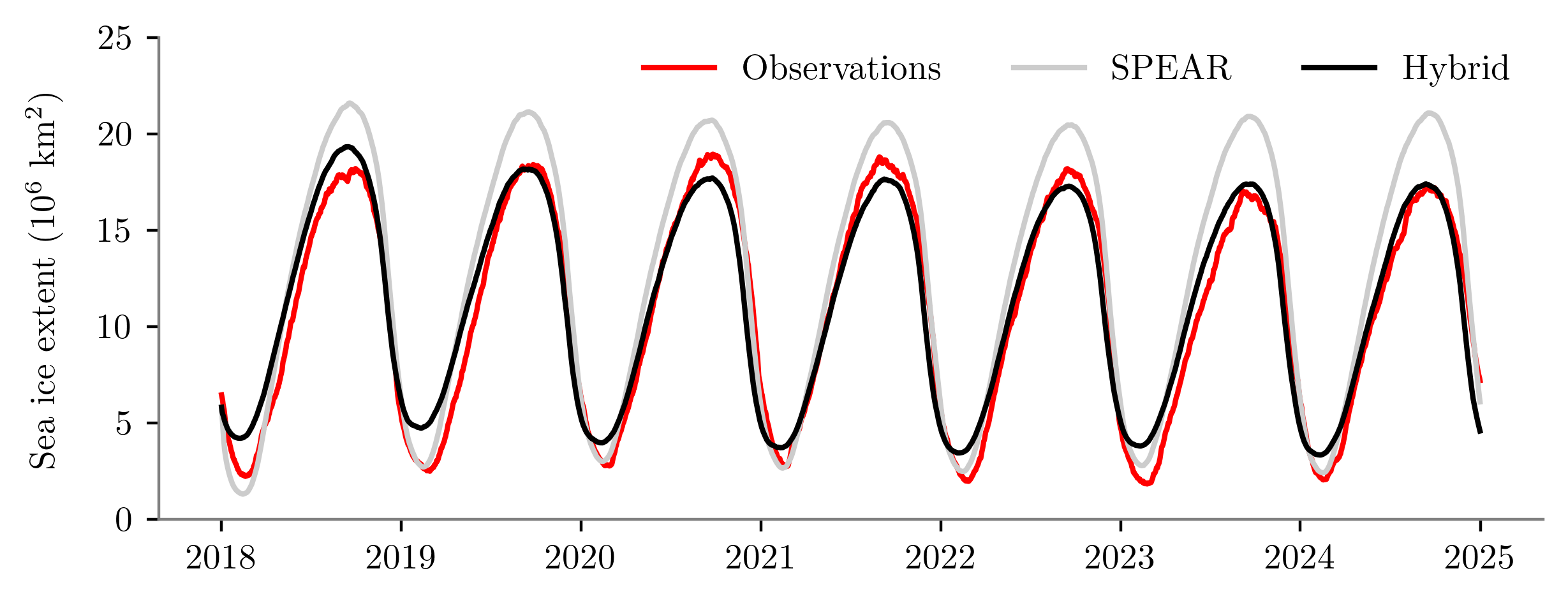} 
	\caption{\textbf{Pan-Antarctic sea ice extent from fully-coupled simulations.}
		A 7-year continuous integration of the fully-coupled GFDL SPEAR climate model (gray) and an equivalent hybrid version of SPEAR (black) which performs online sea ice bias correction using ML.}
	\label{fig:Seaice_coupled} 
\end{figure}

Gregory et al \cite{Gregory2025} demonstrated this phenomenon using the sea ice model error correction approach detailed in the previous section. 
Implementing an ice-ocean trained ML-based sea ice bias correction model into the fully-coupled GFDL SPEAR model during 1-year sea ice forecasts can trigger convective feedbacks in the Southern Ocean. This then preconditions the sea ice and ocean for rapid summertime ice melt through ice-albedo feedbacks. 
Online generalization is significantly improved if the ML model is instead trained on data from a nudged atmosphere configuration of the fully-coupled SPEAR model, highlighting the importance of coupled atmosphere-ice-ocean feedbacks within ML training data. Figure \ref{fig:Seaice_coupled} shows Antarctic sea ice extent from a fully-coupled simulation using the bias-correction approach of Gregory et al \cite{Gregory2025}, but now in a 7-year continuous integration. 
This highlights that the hybrid model can be run stably for multiple years and even shows reduced biases relative to the GFDL SPEAR model. 

\begin{figure}
	\centering
\includegraphics[width=\textwidth]{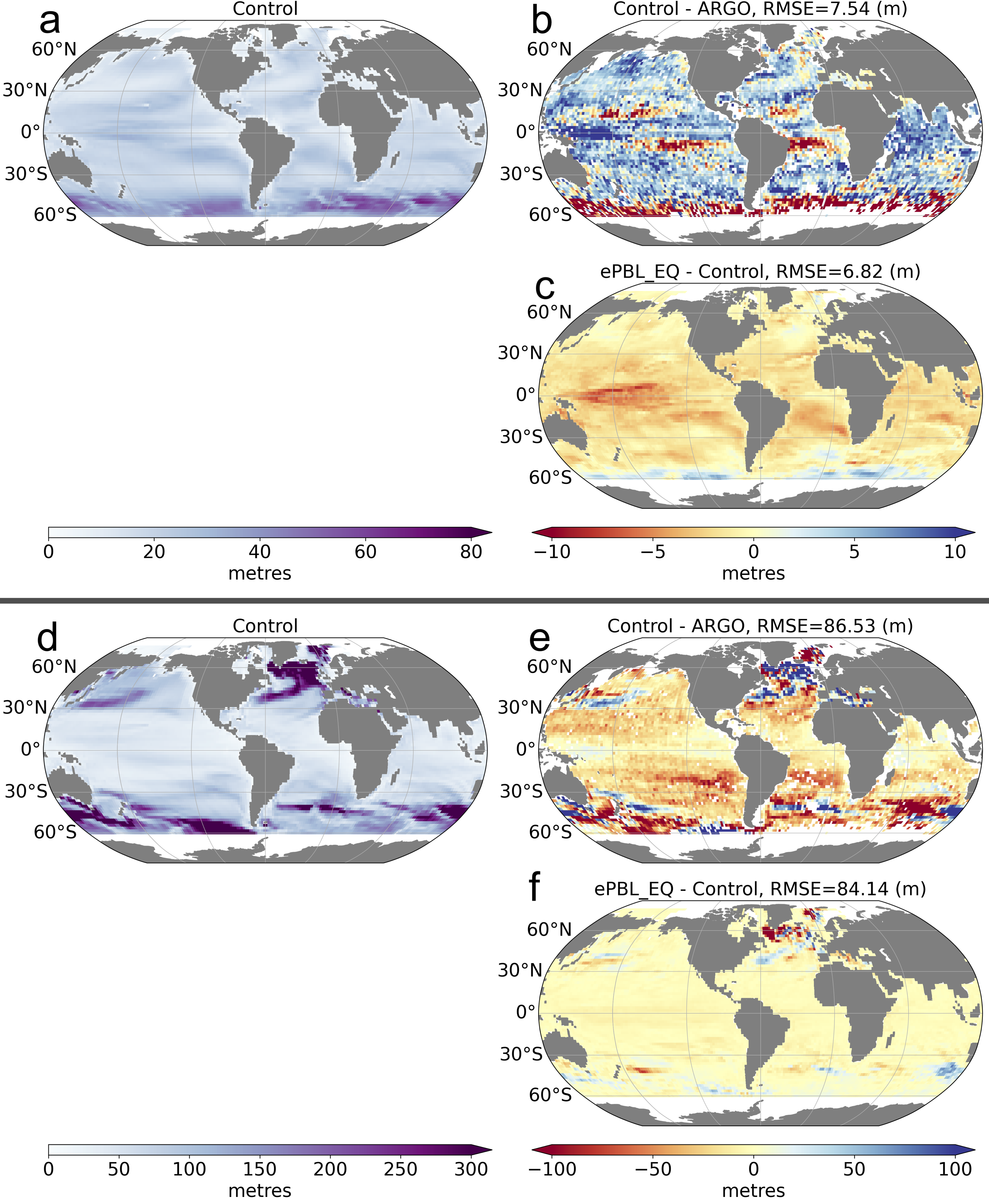} 
	\caption{\textbf{Mixed layer depths from fully-coupled SPEAR  simulations.}
    \textbf{Summer MLD}: (a) Summer MLD averaged over years 1970-2013 in SPEAR run that uses baseline ePBL scheme, (b) Bias in summer MLD between run with baseline ePBL and ARGO data, (c) Difference between ePBL with enhanced diffusivity \cite{Sane2025} and baseline scheme showing the response of the enhancement. Winter MLD: \textbf{d-e-f} are the same as in \textbf{a-b-c}, respectively, but for winter MLD. 
    }
	\label{fig:epbl_EQ_coupled} 
\end{figure}

A similar fully coupled SPEAR model was also implemented using a prototype version of ePBL, incorporating machine-learned diffusivity in equation form \cite{Sane2025}. 
The results of this prototype simulation are encouraging (see Figure \ref{fig:epbl_EQ_coupled}).  The CMIP6 historical model integration runs from 1970--2014 without any stability issues.
This stability is achieved through our approach to implementing generalizable, equation-based diffusivity modifications within a robust ePBL framework, and utilizing data across a wide range of dynamical regimes and forcing conditions encountered by the coupled simulation. 
The impacts of the new scheme on mixed layer depths in this SPEAR prototype model are qualitatively similar to those shown in Figure \ref{fig:mld-summer}. 
Further analysis of the impact on coupled biases is currently underway as part of the development of GFDL's next-generation coupled models. 
However, as discussed previously, improving the model physics does not necessarily lead immediately to a reduction in model bias due to compensating model errors or additional missing physics and feedbacks. 

Therefore, climate models need to be tuned or calibrated. 
The use of simple statistical emulators has been proposed \cite{schneider2017earth} for tuning exercises. However, fully autoregressive AI emulators might also play a role.
Autoregressive AI emulators, trained on reanalysis, can enhance short-term forecast skill at a much cheaper computational cost than traditional numerical weather forecast simulations \cite{bi2023accurate, lam2023learning}.

More recently, AI emulators for climate modeling have emerged to accelerate long-term climate simulations for both the atmosphere \cite{watt2025ace2, chapman2025camulator} and the ocean \cite{dheeshjith2025samudra}. 
These autoregressive emulators are trained on the output of traditional dynamical models. 
These emulators are poised to enhance traditional modeling pipelines in several key ways, which we will describe below. 
First, the low computational cost of emulators enables ensemble sizes that are orders of magnitude larger than those of conventional models, yielding more robust probabilistic forecasts of climate variability. 
Second, they support rapid sensitivity analyses of tuning parameters by allowing repeated, high-fidelity simulations at minimal expense. 
Third, their fast execution makes them ideal as boundary forcings for coupled dynamical simulations. For example, using an ensemble of atmospheric AI emulators to force ocean simulations in OMIP runs can capture a broader range of atmospheric variability than simulations driven by a single reanalysis product. Or by strategically leveraging ML emulators for slow-evolving dynamics while using traditional dynamical models for fast phenomena \cite{meunier2025learning}.
Finally, these emulators are inherently differentiable and are trained on high-fidelity outputs from the traditional model. One can keep the full numerical system ``in the loop'' by using gradient-based optimization and adjoint sensitivity analyses on the original model---opening myriad pathways for deeper scientific insight into Earth system processes or parameter tuning .

Yet, ML autoregressive emulators for climate are still in their early stages of development, and multiple challenges remain to be overcome. 
While traditional models implement operations explicitly derived from physical laws with guaranteed physical consistency, ML emulators may violate fundamental physical principles. 
Incorporating physical constraints into ML architectures, either as hard constraints through architectural design or as soft constraints during training, to ensure outputs respect fundamental laws such as conservation of mass and energy, can help the emulators. 
Similarly to hybrid models, generalization can be challenging to attain due to training on limited subsets of the full climate trajectory space. 
Finally, ML autoregressive emulators for climate appear to be lacking sensitivity to forcing. 
We refer the readers to the most recent papers on this topic \cite{watt2025ace2,dheeshjith2025samudra,chapman2025camulator}.


These approaches aim to combine the computational efficiency of ML methods with the physical consistency and interpretability of dynamical models and methods towards improved representation of multiscale climate dynamics, at lower computational expense. 
The role of AI in climate modeling will likely continue to evolve rapidly, in tandem with more traditional physics-based modeling in the years to come. 

\clearpage 

%
\bibliography{main} 
\bibliographystyle{sciencemag}


\section*{Acknowledgments}
We thank all members of the M$^2$LInES team for helpful discussions and their support throughout this project. 


\paragraph*{Funding:} This research received support through Schmidt Sciences, LLC, under the M$^2$LInES project. This research was also supported in part through the NYU IT High Performance Computing resources, services, and staff expertise.


\paragraph*{Author contributions:}

\noindent 
Conceptualization: LZ, AA, MB, CFG, BR \\
Methodology: LZ, WG, PP, AS, CZ, AA, MB, CFG, BR  \\
Software:  WG, PP, AS, CZ, AA, MB, BR 
Visualization: KAE, RF, DK\\
Validation, Writing – original draft: LZ, WG, PP, AS, CZ, AA, MB, BR, CFG \\
Writing – review \& editing: all authors\\

\paragraph*{Competing interests:}
There are no competing interests to declare.

\paragraph*{Data and materials availability:}
The sea ice model error correction approach has been implemented into the Fortran source code of version 2 of the Sea Ice Simulator (SIS2). Code and documentation for compiling experiments can be found at https://github.com/m2lines/SIS2. Implemented ANN and ZB20 parameterizations can be found at \url{https://github.com/m2lines/ANN-momentum-mesoscale};  for simulation data see \url{https://doi.org/10.5281/zenodo.16058005}. Code for the online implemented GZ21 parameterizations can be found at \url{https://doi.org/10.5281/zenodo.7663074} while the CNN model files used for the online evaluation in this study (GZ21) can be found at \url{https://doi.org/10.5281/zenodo.7663061}; the setup files for the wind-driven double gyre case in GZ21 implementation can be found at \url{https://doi.org/10.5281/zenodo.7663041}.
The figure plotting functions in OM4 configuration can be found at \url{ https://github.com/m2lines/vesri_visuals}. Machine learned vertical mixing enhancements in ePBL are part of NOAA-GFDL MOM6 code package (\url{https://github.com/NOAA-GFDL/MOM6-examples}). 



\subsection*{Supplementary materials}
Figure  S1


\newpage


\renewcommand{\thefigure}{S\arabic{figure}}
\renewcommand{\thetable}{S\arabic{table}}
\renewcommand{\theequation}{S\arabic{equation}}
\renewcommand{\thepage}{S\arabic{page}}
\setcounter{figure}{0}
\setcounter{table}{0}
\setcounter{equation}{0}
\setcounter{page}{1} 


\begin{center}
\section*{Supplementary Materials for\\ \scititle}
\author{
	L.~Zanna$^{1,9\ast}$,\and 
        W.~Gregory$^{2}$,\and
        P.~Perezhogin$^{1}$,\and    
        A.~Sane$^{2}$,\and  
        C.~Zhang$^{2,8}$,\and 
        A.~Adcroft$^{2}$,\and
	M.~Bushuk$^{2}$,\and
        C.~Fernandez-Granda$^{1,9}$,\and
        B.~Reichl$^{2}$,\and        
        D.~Balwada$^{3}$,\and
        J.~Busecke$^{3}$,\and   
        W.~Chapman$^{5}$,\and
        A.~Connolly$^{7}$,\and
        D.~Du$^{2}$,\and
        K.~Everard$^{1}$,\and
        F.~Falasca$^{1}$,\and
        R. Falga$^{1}$, \and
        D.~Kamm$^{6}$,\and
	E. ~Meunier$^{6}$,\and
        Q.~Liu$^{1}$,\and
        A.-A.~Nasser$^{2}$\and
        M.~Pudig$^{1}$\and
        A.~Shao$^{2}$,\and
        J.L.~Simpson$^{7}$,\and
        L.~Vogt$^{1}$,\and
        J.~Wu$^{1}$ \\	
	\small$^{1}$Courant Institute of Mathematical Sciences, New York University, New York 10012, USA.\and
	\small$^{2}$Princeton University Atmospheric and Oceanic Sciences Program, Princeton, USA .\and 
    \small$^{3}$Lamont-Doherty Earth Observatory at Columbia University, Palisades, USA.\and
	\small$^{4}$NOAA Geophysical Fluid Dynamics Laboratory, Princeton, USA.\and
    \small$^{5}$NSF-National Center for Atmospheric Research, Boulder, CO, USA.\and
    \small$^{6}$Sorbonne Université-CNRS-IRD-MNHN, LOCEAN Laboratory, Paris, France.\and 
    \small$^{7}$Columbia University, New York, NY, USA.\and 
    \small$^{8}$Department of Civil and Environmental Engineering, Rowan University, Glassboro, NJ, USA.\and 
    \small$^{9}$Center for Data Science, New York University, New York 10011, USA.\and
	\small$^\ast$Corresponding author. Email: laure.zanna@nyu.edu}
\end{center}

\subsubsection*{This PDF file includes:}
Figure S1






\newpage 

\begin{figure} 
	\centering
  \includegraphics[width=0.95\textwidth]{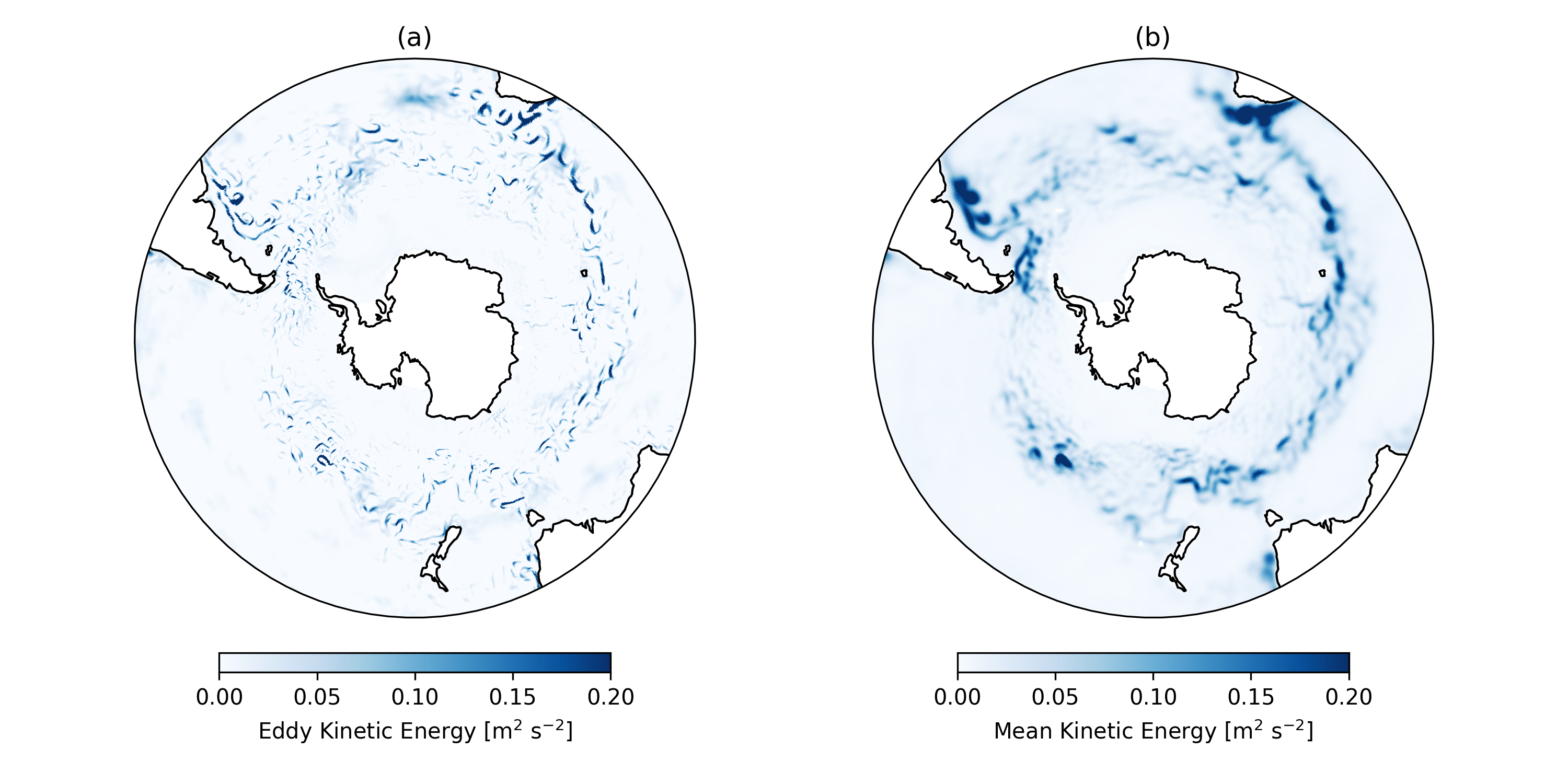}
	\caption{\textbf{Results from a $\frac{1}{4}$-degree, 1-year simulation with a CNN-based parameterization of backscatter.} (A) shows the resolved eddy kinetic energy on December 31 and (B) the mean kinetic energy (averaged over December). The Southern Ocean shown here is one of the regions where we expect a significant amount of energy to be exchanged between the eddy-field and the mean field.}
	\label{fig:gz21_global} 
\end{figure}

\end{document}